# Democratizing Aviation Emissions Estimation: Development of an Open-Source, Data-Driven Methodology

Andy G. Eskenazi[*], Landon G. Butler[+], Arnav P. Joshi[‡], Megan S. Ryerson[◊]

[*]Department of Mechanical Engineering and Applied Mechanics
University of Pennsylvania
Philadelphia, PA

[*]Department of Mathematics
University of Pennsylvania
Philadelphia, PA

[+, ‡,◊]Department of Electrical and Systems Engineering
University of Pennsylvania
Philadelphia, PA

[◊]Department of City and Regional Planning
University of Pennsylvania
Philadelphia, PA

*Corresponding Authors:* andyeske@sas.upenn.edu[*], landonb3@seas.upenn.edu[+], arnavj@seas.upenn.edu[‡]

*Abstract* **— Through an aviation emissions estimation tool that is both publicly-accessible and comprehensive, researchers, planners, and community advocates can help shape a more sustainable and equitable U.S. air transportation system. To this end, we develop an open-source, data-driven methodology to calculate the system-wide emissions of the U.S. domestic civil aviation industry. This process utilizes and integrates six different public datasets provided by the Bureau of Transportation Statistics (BTS), the Federal Aviation Agency (FAA), EUROCONTROL, and the International Civil Aviation Organization (ICAO). At the individual flight level, our approach examines the specific aircraft type, equipped engine, and time in stage of flight to produce a more granular estimate than competing approaches. Enabled by our methodology, we then calculate system-wide emissions, considering four different greenhouse gases ($CO_2$, $NO_x$, CO, HC) during the Landing, Take-off (LTO) and Climb, Cruise, and Descent (CCD). flight cycles. Our results elucidate that emissions on a particular route can vary significantly due to aircraft and engine choice, and that emission rates differ significantly from airline to airline. We also find that $CO_2$ alone is not a sufficient proxy for emissions, as $NO_x$, when converted to its $CO_2$-equivalency, exceeds $CO_2$ during both LTO and CCD.**

## I. Introduction

To combat the rapid warming of Earth's climate, a broad range of stakeholders including planners and policymakers are interested in accurately estimating greenhouse gas emissions from aviation. In response, a multitude of methodologies have been proposed by the research community, which can be broadly categorized into two main groups. The first group are those that are able to calculate system-wide emissions but rely on simplifying heuristics to bypass the nuance of particular aircraft types, engine types, or segmentation of flight [11,12,18,22,23]. The second are those that capture these granularities through more sophisticated methods, but are restricted to particular flights, airports, or regions [27,29]. Furthermore, while there are some methods that are sufficiently detailed and allow for a system-wide calculation, these often use proprietary data and are not open-source [11,14,17], or are highly specialized and thus only accessible to select experts.

Without a methodology that is simultaneously comprehensive, system-wide, and open-source, it becomes difficult to hold decision-makers accountable in the design of an air transportation system [6] that promotes both accessibility [13] and sustainability [1,26,30]. In short, what is needed is an open-source, system-wide emissions estimation model that better captures the nuances of aviation, without coming at the expense of ease of use or limited applicability. The goal of this paper is to provide a method that fulfills these needs.

More precisely, in this work, we build on previous emission estimation methods by leveraging publicly available aircraft, engine, and flight datasets. From this data, we develop a comprehensive procedure for calculating emissions on a per-flight basis, and in-turn, inductively compute both regional and system-wide emissions of the U.S. civil aviation industry.

The contributions of our work are as follows:

- Developing a matching procedure between +200k airplane tail numbers, 816 engine types and 62 airplane types leading to more accurate flight emission estimations using ICAO engine emission factors
- Implementing our procedure to calculate the system-wide CCD and LTO cycle emissions for 1.6 million domestic flights from 2021 Q3 using the Bureau of Transportation Statistic's On-Time dataset
- Incorporating $CO_2$ equivalencies for other greenhouse gases like nitrogen oxides ($NO_x$), carbon monoxide (CO), and hydrocarbons (HC)
- Open-sourcing our data, methodology, and computations in a **publicly available repository**

This work is organized in the following way: a literature review of emissions calculations is presented in Section II; our method to compute the emissions of the air transportation system of the United States is detailed in Section III; the results and analysis from our methods is featured in Section IV; and the final conclusions are delivered in Section V.

## II. LITERATURE REVIEW

### A. System-wide Emissions Calculations

The literature on flight emissions estimation can be classified into two broad categories: those that use simplifying heuristics in order to scale, and those that use sophisticated models but are limited in scope. Across the literature on emissions estimation, there is a lack of open-source tools that can allow researchers, planners, and community advocates to easily estimate emissions associated with aviation.

Many flight emissions models developed in the literature are focused on achieving scalability through rapid estimations; as such, these models avoid detailed flight-level granularity in their emissions calculation methodology and instead employ simplifying assumptions [11,12,18,22,23]. For instance, some previous approaches attempt to calculate the aggregate emissions of an airline network by choosing a representative aircraft or a small sampling of aircraft rather than considering each individual aircraft and engine type [11,12]. In addition, the emissions associated with a flight are attributed to two stages, LTO and CCD (the significance of which is highlighted in the next section of the literature review); some methodologies do not consider these two flight stages separately when calculating emissions [18,22]. Finally, some papers do not incorporate the emissions of greenhouse gasses beyond carbon dioxide such as nitrogen oxides, hydrocarbons, and carbon monoxide [11].

Some papers develop and use more sophisticated models for estimating flight emissions, but their scope is restricted to a particular airport or region. For example, [29] studies the operation times, fuel, and emission parameters of eight emission models during the taxi phase at the Shanghai Hongqiao International Airport. Aircraft emissions during the LTO cycles at the Batumi International Airport in Georgia are calculated using data about aircraft type, engine type, number of passengers, and emissions factors provided by ICAO data in [27]. However, these approaches were not intended to capture system-wide emissions, as would be needed to shape airline network planning decisions. In contrast, Peeters [21], using Europe as a case study, estimates system-wide emissions and demonstrates that the more concentrated a hub-and-spoke network is, the greater is its environmental impact. Morrel and Lu [18] build on Peeters' work to find that hub by-pass networks produce much lower greenhouse gas emissions and noise pollution compared to the hub-and-spoke networks.

There are some emissions estimation methods that are both sophisticated and are suitable for a system-wide calculation of emissions. Miyoshi and Mason [17] use data about aircraft types and flight stages to calculate emissions for 1,626 routes in the domestic UK, intra-EU, and North Atlantic markets. The FAA developed the SAGE model [11,14] that is detailed enough to not only model emissions at the level of a single flight, but is capable of analyses at the regional, national, and global levels. However, the datasets used in these methods are often proprietary and require a high level of technical expertise to understand them, making these tools inaccessible for most individuals.

We fill a gap in the aviation emissions calculation literature by proposing an approach for calculating emissions that uses publicly accessible data and methods, models emissions at the level of a flight, and can be used to calculate system-wide emissions. The next section emphasizes the importance of the segmentation of flight stages in the process of computing system-wide greenhouse gas emissions.

### B. Segmentation of Flight Stages

Airplane engine emissions vary at each stage of flight, given that engines operate at different thrust settings in each stage. In an attempt to characterize and certify jet engines, the International Civil Aviation Organization (ICAO) has defined two standardized cycles in which an airliner's flight can be divided into, most notably Landing and Take-off (LTO) and Cruise, Climb and Descent (CCD) [28]. The first of these cycles LTO, is used to denote all flight activities that take place up to 3,000 ft. from ground-level, while the second, CCD, characterizes activities beyond 3,000 ft. Each of these phases can in turn be split into different stages, as Figure 1 shows.

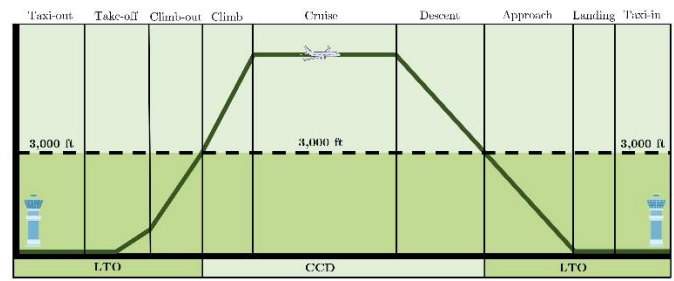

Figure 1.   Landing and Take-off (LTO) and Cruise, Climb and Descent (CCD) flight cycles.

TABLE I. ICAO'S THRUST AND TIME SETTINGS FOR LTO CYCLE

| Operating Mode | Thrust Setting | Standard Time (*minutes*) |
|---|---|---|
| Take-off (*abbr.* T/O) | 100% | 0.7 (42 *s*) |
| Climb-out (*abbr.* C/O) | 85% | 2.2 (132 *s*) |
| Approach-landing (*abbr.* APP) | 30% | 4 (240 *s*) |
| Taxi/ground idle (*abbr.* IDLE) | 7% | 26 (1560 *s*) |

According to the International Air Transport Association (IATA), LTO cycles contribute around 10% of all emissions in a flight, accounting for the highest ratio of emissions per distance flown [8]. For this reason, short-haul flights tend to have a much stronger environmental impact than long-haul flights due to the LTO cycle representing a higher percentage of total flight time [9]. Additionally, LTO cycles can compose up to 70% of an airport's total emissions, and 30% of a flight's entire CO and HC emissions [20]. The standard way in which the engine thrust settings of an LTO cycle are defined, according to the ICAO, are shown in Table I, alongside the operation time of each setting.

## III. METHODOLOGY

### A. Conceptual Framework and Introductory Example

The methodology that we developed to estimate the system-wide flight emissions from the domestic route network of the United States merges phase of flight information with detailed engine and aircraft characteristics to capture the emissions of four different greenhouse gases across LTO and CCD. To achieve the implementation of this, we utilized six different, publicly available data tables provided by the Bureau of Transportation Statistics (BTS), the Federal Aviation Administration (FAA,) EUROCONTROL, and ICAO. The global picture of how each data source is related and utilized can be seen in Figure 2.

To motivate this section and how our methodology uses and integrates these data tables, let us examine the following computational example. Consider flight DL2441, a Delta Airlines flight operating between Philadelphia and Atlanta. A frequent operator of this route is the airliner tail-numbered N815DN, a B737-900ER that seats 180 passengers, according to data from the BTS Schedule B-43 Inventory (also known as the Aircraft by Airline database). Furthermore, according to the FAA's Tail Registry and the ICAO's Aircraft Engine Emissions Databank [10], this aircraft is equipped with two CFM56-7B27E engines. Looking at the BTS's On-Time Flights Schedule Statistics, we find that on average, DL2441 has an air time of 124 minutes, and N815DN in particular averages 7.43 and 15.42 minutes during taxi-in and taxi-out, respectively, resulting in a total of 22.85 minutes of average idle time. With all of this data, it is possible to compute the emissions released by flight DL2441.

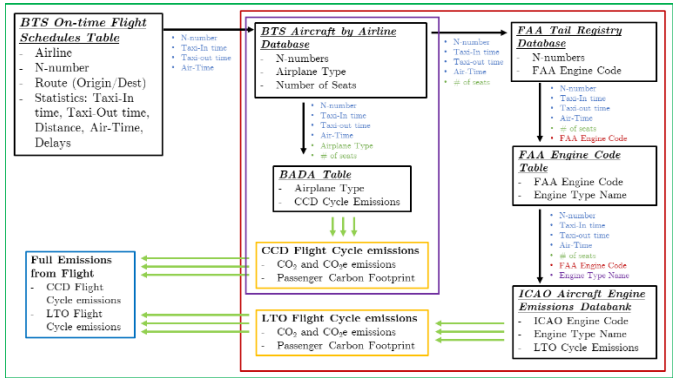

Figure 2. Data Matching - Each table is underlined, while the variables coming from each table are highlighted under each arrow.

TABLE II. EXAMPLE EMISSIONS FACTORS FOR LTO CYCLE

| Engine Type: CFM56-7B27E | | | | |
|---|---|---|---|---|
| Stage of Flight | HC (kg/s) | $CO_2$ (kg/s) | CO (kg/s) | $NO_x$ (kg/s) |
| Take-off | 0.00003879 | 4.07295 | 0.00040083 | 0.03095442 |
| Climb | 0.00002062 | 3.24765 | 0.00017527 | 0.01844459 |
| Approach | 0.00001715 | 1.08045 | 0.00096726 | 0.00311787 |
| Idle | 0.0001694 | 0.3465 | 0.0032329 | 0.0004796 |

Using the ICAO's Aircraft Engine Emissions Databank [10], we can extract the engine emissions factors for greenhouse gases like $CO_2$, CO, $NO_x$, and HC, as shown in Table II. Then, using the standard times listed in Table I above, in addition to the idle time reported in the previous paragraph, it is possible to calculate the emissions from the LTO flight cycle. In this case, the total amount of HC, $CO_2$, CO and $NO_x$ produced during the LTO cycle of this flight is equal to 0.24, 1334.11, 4.70 and 5.14 kg respectively, or 2893.61 kg of $CO_2$e (when converting HC, CO and $NO_x$ into $CO_2$ terms using the conversion factors reported by [5]).

Next, using EUROCONTROL's Base of Aircraft Data (BADA) [7], it is possible to extract the flight emissions factors for the same greenhouse gases but corresponding to the CCD flight cycle. This is shown in Table III below:

TABLE III. EXAMPLE EMISSIONS FACTORS FOR CCD CYCLE

| Airplane Type: B737-900ER | | | | |
|---|---|---|---|---|
| Duration (min) | HC (kg) | CO2 (kg) | CO (kg) | NOx (kg) |
| 22 | 0.35 | 3114 | 2.93 | 18.3 |
| 32 | 0.51 | 4626 | 4.05 | 27.47 |
| 39 | 0.57 | 5608 | 4.36 | 32 |
| 71 | 0.84 | 10147 | 5.64 | 52.69 |
| 105 | 1.13 | 14300 | 7.12 | 70.1 |
| 139 | 1.44 | 18294 | 8.26 | 86.64 |
| 206 | 1.97 | 26953 | 10.42 | 123.27 |
| 273 | 2.5 | 36023 | 12.62 | 162.82 |
| 340 | 3.03 | 44475 | 14.73 | 197.69 |
| 410 | 3.55 | 54250 | 17.11 | 240.25 |

As can be seen, each emission factor is associated with a different duration of flight. Thus, knowing that DL2441 on average spends 124 minutes in the sky, it is possible to perform linear interpolation and find the emissions corresponding to the CCD cycle. For this flight and cycle, the total amount of HC, $CO_2$, CO and $NO_x$ produced is equal to 1.31, 16,564.99, 7.77 and 79.48 kg respectively, or 40,371.85 kg of $CO_2$e. Adding the contributions from the LTO and CCD flight cycle together, what this means is that DL2441's total environmental impact, induced by the CFM56-7B27E-equipped B737-900ER, is 43,265.46 kg of $CO_2$e, resulting in a footprint of 240.36 kg of $CO_2$e per passenger.

This example details the level of granularity with which we decided to calculate the system-wide emissions of the domestic airline industry in the United States. Our novel approach is to look at each flight considering not only the airplane type operating that route, but also the engine type it is equipped with. Furthermore, we also incorporate real taxi-in and taxi-out times, as to return the most accurate results. This process is explained in greater detail in the remaining parts of this section.

### B. Engine Emission Factors

At the heart of our emissions calculator lies ICAO's Aircraft Engine Emissions Databank [10] and EUROCONTROL's Base of Aircraft Data (BADA) [7]. The first of these tables agglomerates engine specifications for more than 800 different models, collected from manufacturers certifying their engines against the ICAO's emission standards. Currently, ICAO has set standards for greenhouse gases like $CO_2$, CO, $NO_x$, and HC, and as such, their engine databank is composed of rich information regarding the emissions of these gases during the landing and take-off (LTO) flight cycle. Thus, for a particular engine type, it is possible to extract emissions factors for take-off (T/O), climb-out (C/O), approach-landing (APP) and taxi/ground idle (IDLE) for a particular pollutant, and calculate the corresponding emissions from the LTO cycle. Taking $CO_2$ as an example, we find that the total emissions from LTO are:

$$E_{CO_2,LTO} = E_{CO_2,T/o}t_{T/O} + E_{CO_2,c/o}t_{c/o} + E_{CO_2,APP}t_{APP} + E_{CO_2,IDLE}t_{IDLE} \qquad (1)$$

$E_{CO_2,T/O}$ denotes the emissions factor (in Kg/s) for $CO_2$ during take-off, and $t_{T/O}$ the corresponding standard LTO time (as provided in Table 1). This simple equation can be repeated for CO, $NO_x$, and HC, and the emissions can be added together by taking their equivalent effect in $CO_2$ as follows (using the conversion factors reported by [5]):

$$E_{CO_2e,LTO} = E_{CO_2,LTO} + 1.57 * E_{CO,LTO} + 84 * E_{HC,LTO} + 298 * E_{NO_X,LTO} \qquad (2)$$

To calculate emissions for the remaining phase of flight, we utilize EUROCONTROL's Base of Aircraft Data (BADA) [7], since it contains valuable information surrounding the emissions from the cruise, climb and descent (CCD) flight cycle. In essence, the table provides altitude and distance-dependent emissions data for most commercial aircraft in operation today in the United States. However, because of the complexities that arise due to changes in atmospheric properties at different altitudes, in addition to the difficulties in modelling weather patterns, the data in BADA's tables is tabulated for a specific set of cases. In other words, for each of the reported airplane types, the emissions for each greenhouse gas are computed for a pre-designated set of flight distances (or flight times), ranging from 125 nautical miles up to 8000 nautical miles. To compute the CCD emissions for a particular flight, it is necessary to perform linear interpolation, since it is highly likely that a flight's distance or time will fall in between two of the tabulated cases in the table (as all domestic U.S. flights do). Taking $CO_2$ once again as an example, we find that the emissions from the CCD cycle can be computed as:

$$E_{CO_2,CCD} =$$
$$\frac{\lceil E \rceil_{CO_2,CCD} - \lfloor E \rfloor_{CO_2,CCD}}{\lceil d \rceil_{CCD} - \lfloor d \rfloor_{CCD}}(d - \lfloor d \rfloor_{CCD}) + \lfloor E \rfloor_{CO_2,CCD} \qquad (3)$$

In this equation, $d$ denotes a flight's distance, $\lfloor d \rfloor_{CCD}$ is the lower bound from the tabulated list of distances in which $d$ lies, and $\lceil d \rceil_{CCD}$ is the upper bound, while $\lceil E \rceil_{CO_2,CCD}$ and $\lfloor E \rfloor_{CO_2,CCD}$ represent the calculated $CO_2$ for CCD corresponding to the upper and lower bound flight distance. Due to the complexities mentioned previously, the estimate of the emissions obtained from the CCD flight cycle is less accurate than those for the LTO. Furthermore, the tabulated emissions factors from BADA are airplane specific and not engine specific as was the case with the ICAO's engine emissions databank. As a result, whether a 737-800 is fitted with a newer or older type of engine does not matter in CCD terms since it would result in the same emissions. Nonetheless, despite these setbacks, this method for calculating emissions is still more accurate than most other methods used in the literature, especially compared to those who do not distinguish between airplane types. It should be noted, similarly to LTO, the carbon equivalent emissions from the CCD cycle can be found by multiplying each greenhouse gas by its respective conversion factor, just like in equation (2).

Finally, to find a flight's total emissions, the emissions from LTO and CCD can be added, resulting in:

$$E_{CO_2e} = E_{CO_2e,LTO} + E_{CO_2e,CCD} \qquad (4)$$

*C.    Per-Flight Estimations*

The innovation in our work lies in using our calculator to compute the emissions for every domestic flight in the United States. However, to accomplish this, we need to access some essential information about each flight, including the airplane type used, the equipped engine, the number of seats, the airtime, as well as the taxi-in and taxi-out times (which would allow us to even more accurately compute LTO emissions). One of the data sources that contains much of this information is the United States' Bureau of Transportation Statistics (BTS) On-Time flight schedules. The BTS tables provided us with valuable monthly historic flight data of the United States civil aviation market beginning in January 1987. For the purposes of our work, we consider July, August, and September of 2021, as this was the most recent three-month span at the time of our analysis. Historically, September has been one of the strongest months in terms of domestic civil aviation activity in the United States [6], so we deemed it to be an appropriate choice. Each one of the +500k rows in the BTS On-Time flight schedules table denotes a unique flight, and its entries include the operating airline, the airplane N-number, the route origin and destination airport, as well as some key statistics like taxi-in time, taxi-out time, distance flown, airtime, and delays. However, this data does not have information about the airplane type used, its engine type, and its seating capacity.

Since the BTS data contains the N-number, we can cross-reference this number with BTS's Schedule B-43 Inventory, which contains +100k N-numbers in the United States and provides access to the corresponding airplane type and number of seats. Within Schedule B-43, an airplane type is often referred to using many different notations (such as B737-800, 737/800 or 737-8NG); thus, it is important to clean and standardize the notation to produce a unique airplane identification name. Schedule B-43, together with the On-Time flight schedules tables, provides the information needed to compute the emissions from the CCD flight cycle with the BADA dataset. However, in order to calculate the LTO emissions, it is important to find the corresponding engine type that was used on a given route of interest.

We can find this information in FAA's Tail Registry Database, which for every tail-number in the United States, contains the associated engine type used. However, we the FAA's notation for engine types often deviates significantly from that reported by the ICAO (known as unique engine ID) in their aircraft emissions databank. Using the FAA's Engine Code table, we can run a string token matching procedure (Jaccard Similarity) to match the FAA's engines to those of the ICAO. The results from this notation translation process successfully pair most of the engines from each table; for the cases in which our procedure couldn't find a match, we assigned the most popular engine type for the particular type of airplane (for instance, for the 737-800, this would be the CFM56-7B27).

Additionally, it should be mentioned that whenever data for a particular airplane in an airplane family is not available (such as for the 737-8, 737-9, A320NEO, A330-900, or A220), we compute the flight emissions using data from the closest family member (such as 737-800, 737-900, A320-200, A330-300 and A320-200 respectively) and adjust the result by an efficiency savings percentage reported by the manufacturer [3,4,15].

TABLE IV. AIRLINE SUMMARY STATISTICS FOR 2021 Q3

| Airline Statistics Summary – 2021 Q3 | | | | | | | | |
|---|---|---|---|---|---|---|---|---|
| Airline | | Total Flights | Emission Flights | Number of Seats | Total $CO_2$ (kg) | $CO_2$ (kg) per Seat Mile | Total $CO_2e$ (kg) | $CO_2e$ (kg) per Seat Mile |
| Name | Code | | | | | | | |
| Southwest | WN | 305,916 | 285,741 | 41,918,850 | 3.830E+09 | 0.126 | 9.080E+09 | 0.305 |
| American | AA | 213,996 | 204,603 | 32,752,807 | 4.130E+09 | 0.131 | 1.040E+10 | 0.329 |
| SkyWest | OO | 210,714 | 206,516 | 13,676,292 | 1.320E+09 | 0.220 | 2.640E+09 | 0.447 |
| Delta | DL | 208,252 | 198,876 | 32,803,502 | 3.860E+09 | 0.128 | 9.100E+09 | 0.308 |
| United | UA | 132,634 | 123,622 | 20,040,396 | 3.060E+09 | 0.126 | 7.470E+09 | 0.306 |
| Republic | YX | 94,309 | 87,877 | 6,747,929 | 5.680E+08 | 0.169 | 1.120E+09 | 0.337 |
| Envoy | MQ | 70,066 | 68,249 | 4,403,904 | 3.520E+08 | 0.189 | 7.160E+08 | 0.399 |
| Endeavor | 9E | 70,063 | 69,648 | 4,707,932 | 3.770E+08 | 0.243 | 7.620E+08 | 0.498 |
| PSA | OH | 62,268 | 59,970 | 4,801,970 | 3.570E+08 | 0.203 | 7.180E+08 | 0.412 |
| JetBlue | B6 | 54,956 | 55,812 | 7,800,411 | 1.210E+09 | 0.132 | 2.900E+09 | 0.301 |
| Spirit | NK | 54,956 | 47,624 | 8,764,933 | 8.320E+08 | 0.099 | 2.020E+09 | 0.244 |
| Alaska | AS | 52,698 | 50,878 | 8,397,814 | 1.110E+09 | 0.120 | 2.620E+09 | 0.291 |
| Mesa | YV | 45,166 | 42,627 | 3,322,779 | 3.180E+08 | 0.160 | 6.300E+08 | 0.321 |
| Frontier | F9 | 39,019 | 34,462 | 6,604,642 | 5.690E+08 | 0.097 | 1.390E+09 | 0.240 |
| Hawaiian | HA | 18,965 | 18,677 | 2,972,301 | 4.360E+08 | 0.182 | 1.040E+09 | 0.409 |

Applications of this methodology are limited to flights within the U.S. civil aviation system, as this is the only scope with publicly available flight, aircraft, and engine emission factor data. We also do not explicitly account for per-flight atmospheric conditions which influence necessary thrust and thus emissions, as was done in [25,31]. However, adverse weather conditions generally result in longer flight times, which is captured by our methodology. Lastly, we do not scale emissions at altitude to account for higher radiative forcing and global warming potential [16,22]. Though, because our methodology already segregates emissions into the LTO and CCD flight stages, further work could apply scaling factors to account for aviation induced radiative forcing.

## IV. RESULTS AND ANALYSIS

Using the methodology presented in the previous section and BTS's On-Time dataset, we examined ~1.65 million flights corresponding to the third quarter (July, August, September) of 2021 in the domestic US civil aviation network. Of these, we were able to compute the greenhouse gas emissions for ~1.55 million flights, or 93.8% of the system's flights (due to missing data on BTS's On-Time dataset). Overall, the results from this system-wide calculation can be quickly recreated using the code available in our repository.

The data we obtained is incredibly rich, given that for every flight in the United States for 2021 Q3, we were able to compute both their LTO and CCD emissions, with the LTO split between the origin and destination airports. Furthermore, we reported both the total $CO_2$ and $CO_2e$ emissions of each flight, as well as the individual contributions of CO, HC and $NO_x$ in both the LTO and CCD flight stages. In addition, for each flight, we obtained information regarding the particular airplane type and the operating carrier. As a result, it is possible to aggregate this data across various fields, such as the operating carrier, origin/destination airport, type of greenhouse gas, route, airplane type, and engine type.

As an example, on grouping the data by the operating carrier, it is possible to see the environmental impact induced by each airline during 2021 Q3. This can be seen in Table IV.

Using the $CO_2$ per seat-mile ratio as a metric for comparison, we found that Frontier, Spirit, Alaska, United and Southwest rank lowest. The corresponding values we found were 0.097, 0.099, 0.120, 0.126 and 0.126 kg of $CO_2$ per seat-mile respectively, which are close to those reported by a similar, less granular analysis performed in [2] and a study of fuel consumption led by Bo Zou [32].

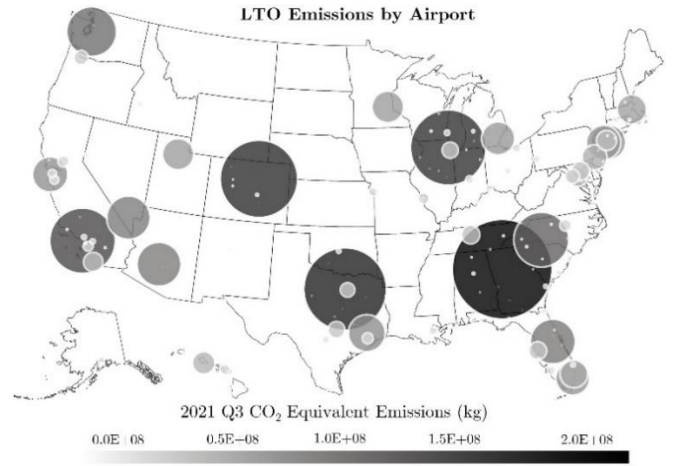

Figure 3. Local LTO emissions produced at every airport of the United States during 2021 Q3.

Equally interesting results were obtained when grouping emissions by airport. In particular, by summing up the greenhouse gas emissions from the LTO cycles, we determine the local effect induced by civil aviation at each U.S. airport as shown in Figure 3.

As mentioned in the literature review, LTO cycles have the strongest regional environmental effect, since they occur at an altitude below 3000 ft. We find that the top five polluting airports are Atlanta Hartsfield-Jackson (ATL), Dallas/Fort Worth (DFW), Denver (DEN), Chicago O'Hare (ORD) and Los Angeles (LAX) with emissions during 2021 Q3 equal to $2.22*10^8$, $1.84*10^8$, $1.73*10^8$, $1.69*10^8$ and $1.46*10^8$ kg of $CO_2e$, respectively. The sheer magnitude of these emissions has tremendous proven consequences on human health and the environment [9], which have been monetized by [19]. For the mentioned top-ranking airports, the economic impacts range from $50 to $200 million USD per year.

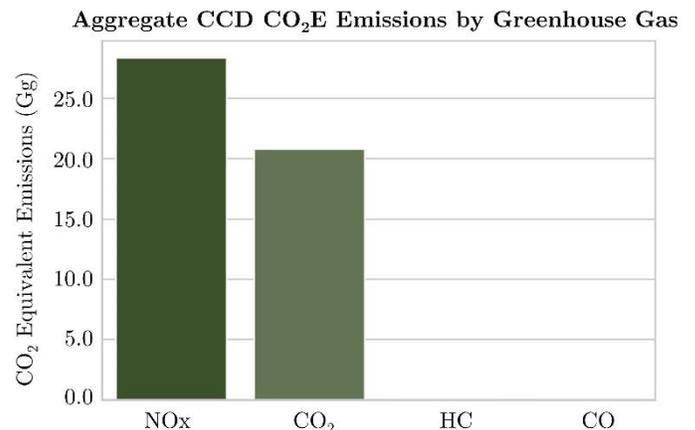

Figure 4. Environmental effects of $NO_x$, HC and CO in terms of $CO_2$, for the LTO flight cycle.

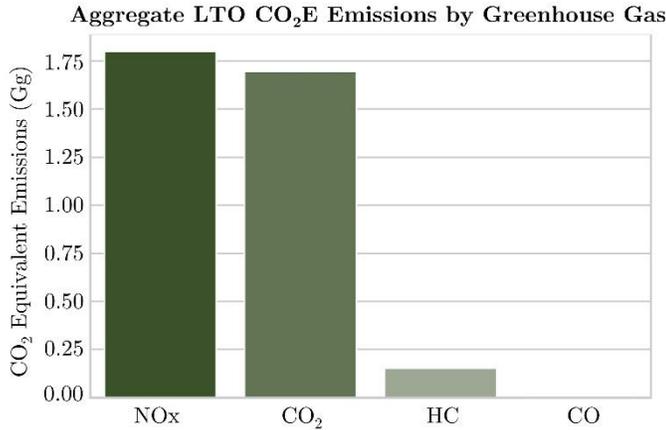

Figure 5.  Environmental effects of NOx, HC and CO in terms of CO₂, for the CCD flight cycle.

To further our understanding of the environmental impact in the regions surrounding airports, we also analyzed the individual contribution of each greenhouse gas during LTO. This can be seen in Figure 4, which contains the emissions calculated for NOx, CO and HC converted to their equivalent effects in $CO_2$ terms for LTO using the conversion factors reported by [5]. As a point of comparison, Figure 5 provides the same calculations for CCD. As is evident in these two figures, the $CO_2$ equivalent effect of NOx exceeds that of $CO_2$ alone in both flight stages, despite having a much smaller mass generated during flight.

All in all, the data that our method generated has allowed us to visualize air transportation emissions from a system-wide perspective, incorporating the effects of the primary greenhouse gases. To conclude our discussion of the results section, below we present two figures that greatly illustrate flight emissions macroscopically.

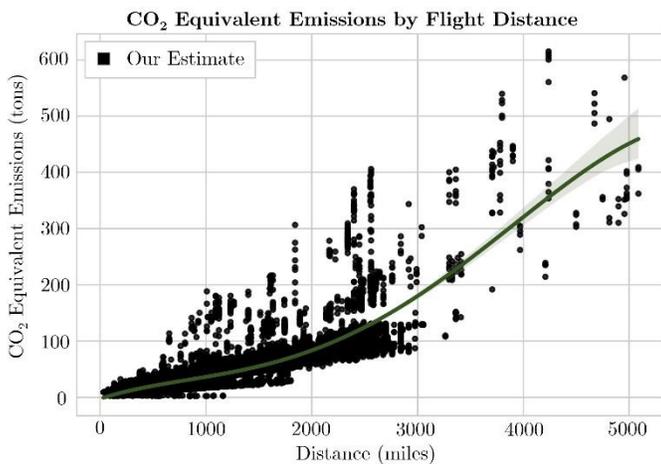

Figure 6.  CO₂e emissions against the flight distance for 2021 Q3.

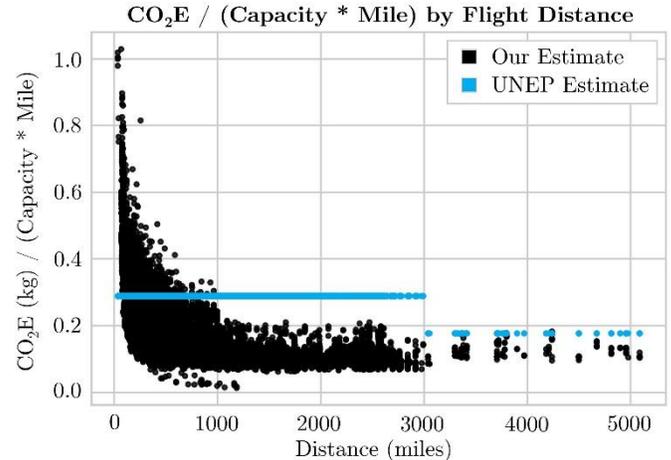

Figure 7.  CO₂ emissions per seat-mile against flight distance for 2021 Q3.

Figure 6 plots the CO₂e emissions against flight distance. An interesting pattern to highlight is the presence of varying emission estimates for the same flight distance. This pattern represents flights which travel the same distance, but historically have utilized multiple airplane types with different engine emission factors. Another interesting phenomenon shown in Figure 6 is the clear divide for flights shorter or longer than 3000 miles. This divide exists because short distance flights utilize range-limited aircraft types such as the B737 and the A320, whereas long distance flights utilize aircraft types such as the A330, B777 and B787.

A variation of Figure 6 is shown in Figure 7, which instead shows the relationship between $CO_2$ emissions per seat-mile and flight distance. Additionally, we plot the estimates following the methodology provided by the United Nations Environmental Program for $CO_2$ emissions per seat-mile for short and long-haul flights [22]. As can be seen, our results provide a much more comprehensive picture than the values reported by UNEP. Our computations illustrate that incorporating aircraft type, engine type, and flight segmentation provide a much more nuanced emissions estimate compared to using constant emissions for short and long-haul flights. Another point worth highlighting is that short haul flights carrying few passengers have the highest emissions per seat mile ratio, as the rapidly decaying curve on the left side of the plot illustrates. However, past 1000 miles, our estimates seem to stabilize into a constant ratio.

## V. CONCLUSIONS

This research proposes a novel method for estimating per-flight emissions for four different greenhouse gasses by considering aircraft, engine, and flight stages; when applied to each flight in BTS' On-Time dataset from 2021Q3, we gain insights into the system-wide emissions from the U.S domestic airline industry. We find that for the same distance traveled, aircraft and engine choice contribute to significant heterogeneity in emissions. In turn, airlines produce vastly different emissions per seat mile influenced by their choice of active fleet. Our calculation also highlights the considerable amount of $NO_x$ emissions in terms of $CO_2$-equivalency in both LTO and CCD.

With the development of hydrogen airplanes and biofuels, our methodology can help stakeholders model how these sustainable aviation technologies can drive the industry towards net zero [1]. Due to the wealth of publicly-available transportation data, we advocate for the continued use of data-driven methodologies in shaping policy and network planning decisions.